\documentstyle[12pt]{article}

\begin{document}

\title{About Black Holes Without Trapping Interior }
\author{Alejandro Cabo and Eloy Ay\'on \\
\\
{\it Instituto de Fisica, Universidad de Guanajuato,}\\
{\it Loma del Bosque, No. 103, Frac. Lomas del Campestre,}\\
{\it Le\'on, Guanajuato, M\'exico,} \\
\\
{\it and }\\
\\
{\it Grupo de Fisica Te\'orica, ICIMAF}\\
{\it Calle E, $\sharp$ 309, Vedado}\\
{\it Habana 4, Cuba.}}
\date{}
\maketitle

\begin{abstract}
\noindent 
Physical arguments related with the existence of black holes solutions
having a nontrapping interior are discussed. Massive scalar fields
interacting with gravity are considered. Interior asymptotic solutions
showing a scalar field approaching a constant value at the horizon are
given. It is argued that the coupled Einstein-Klein-Gordon equations can be
satisfied in the sense of the generalized functions after removing a
particular regularization designed for matching the interior solution with
an external Schwartzschild spacetime. The scalar field appears as just
avoiding the appearance of closed trapped surfaces while coming from the
exterior region. It also follows that the usual space integral over $T_0^0$
in the internal region just gives the total proper mass associated to the
external Schwartzschild solution, as it should be expected.
\end{abstract}

\pagestyle{empty}

\newpage

\section{Introduction}

The question about the final states of collapsing massive stars is a central
issue in Astrophysics. Black holes are expected to be the ultimate
configurations of stars being merely twenty times more massive than the Sun 
\cite{ellis1}. The general point of view asserts that after the collapse a
central singularity develops which is surrounded by an interior region being
causally isolated from outside\cite{wein}.

That is the structure shown by all the known black hole solutions, and which
also is strongly suggested by the so called singularity theorems \cite{hawk}%
. It is a fact that in all the standard solutions, it is possible to find
trapped surfaces at arbitrarily near distances from the horizon from the
inside. However, it is not evident that all the physical solutions should
have such a behavior.

The aim of this work is to discuss physical considerations related with the
existence of black hole solutions having a ''normal'' interior space-time
without closed trapped surfaces \cite{hawk}. The Lagrangian system seeming
to allow configurations being of interest in this sense corresponds to the
massive scalar field interacting with gravity. Here we will consider a
solution being spherically symmetric. It shows a scalar field tending to a
constant at some ''assumed'' horizon in such a way that this horizon can be
approached from the interior.

The plan of the work goes as follows.

In first place, a regular asymptotic solution of the Einstein and
Klein-Gordon equations is obtained in the neighborhood of the origin and
numerically extended to a probable horizon at some radius $\rho_0$. The
solution has a scalar field which increases up to a finite value at $\rho_0$%
. The asymptotic field configuration was also discussed in the related work 
\cite{eloy} in the preparation of which we started to investigate the
problems addressed here.

The analytic behavior corresponding to the considered numerical solution in
the neighborhood of the expected horizon is also determined. Afterwards, the
question of matching the Schwartzschild solution at the exterior is
considered. For this purpose the singular field configuration consisting of
the considered solution at the interior and the Schwartzschild one at the
exterior, is regularized. The scalar field is assumed as vanishing in the
outside zone in accordance with the no-hair theorems. Therefore, the scalar
field shows a rapid variation near the horizon. The inverse of the radial
component of the metric tensor near the horizon, tends to vanish on both
sides. This property allows to design a regularization which can smooth out
the strong singularities of the scalar field derivative in the kinetic
energy terms of the Einstein equations. Effectively, after properly
selecting the regularization for the three fields involved, it can be argued
that they solve Einstein-Klein-Gordon equations in the sense of the
generalized functions.

It also follows that the matching conditions fully determine all the
parameters fixing the starting asymptotic solution in terms of the
total black hole mass. In addition, the  contribution of the internal
scalar fields to the integral determining the total  proper mass, exactly 
coincides with the mass associated to the external Schwartzschild solution.

The possibility for the exchanging of real probe particles across the
horizon is also conjectured after remarking that: 1) Any arbitrarily small
(but real) probe particle would make a nonvanishing back-reaction on the
metric being able to locally disrupt the horizon surface when the particle
is sufficiently near it. 2) Outgoing particle trajectories of the
Schwartzschild space-time always exist for arbitrarily small but finite
distances from the horizon from the outside. 3) If the back-reaction is able
to allow a falling from the inside probe particle to pass out just a little,
then it could go far away the horizon following an outgoing particle
trajectory. This can be so because the particle accumulated sufficient
kinetic energy during the falling out at the interior.

In Section 2 the interior solution is discussed. Section 3 is devoted to
discuss the matching with the Schwartzschild space-time. Finally in a final
section conclusions and possibilities for the continuation of the work are
given.

\section{The interior solution}

Let us consider the coupled set of Klein-Gordon and Einstein equations in
spherical coordinates. It can be reduced to the form

\begin{eqnarray}
{\frac{{u^{\prime }(\rho )}}\rho }-{\frac{{1-u(\rho )}}{\rho {^2}}} &=&{}-{(}%
8\pi k/c^4)({\frac{{u(\rho )\,}\varphi {{{^{\prime }(\rho )}^2}}}2}+{\frac{%
{\it m}^2\ \varphi {{{(\rho )}^2}}}2)} \\
{\frac{u(\rho )}{v(\rho )}}\,{\frac{{v^{\prime }(\rho )}}\rho }-{\frac{{%
1-u(\rho )}}{\rho {^2}}} &=&{}+{(}8\pi k/c^4)({\frac{{u(\rho )\,}\varphi {{{%
^{\prime }(\rho )}^2}}}2}-{\frac{{\it m}^2{\it \ }\varphi {{{(\rho )}^2}}}2)}
\\
m^2\varphi (\rho )-u(\rho )\,\varphi ^{\prime \prime }(\rho ) &=&\varphi
^{\prime }(\rho )\,\left( {\frac{(u(\rho )+1)}{\rho u(\rho )}}- {(}8\pi
k/c^4){\frac{\rho {\it m}^2\varphi {(\rho )}^2}{2u(\rho )}}\right)  
\end{eqnarray}

\noindent 
where the invariant interval has been taken as defined by the spherical
coordinates form

\[
ds^2={\it v}(\rho ){dx^o}^2-u(\rho )^{-1}d\rho ^2-\rho ^2(sin^2\theta d\phi
^2+d\theta ^2). 
\]

The scalar field $\phi $ is real and its mass $m$ can be absorbed in the
definition of a new radial variable $r=m\rho $. Moreover, the scalar field
also can been scaled as $\varphi =\phi /\sqrt{8\pi k/c^4\ }$in order to
absorb  the factor $8\pi k/c^4$ multiplying the energy momentum
tensor in it. After these changes, the equations take the form

\begin{eqnarray}
{\frac{{u^{\prime }(r)}}r}-{\frac{{1-u(r)}}{{r^2}}} &=&{}-{\frac{{u(r)\,\phi 
{{^{\prime }(r)}^2}}}2}-{\frac{\phi {{{(r)}^2}}}2} \\
{\frac{u(r)}{v(r)}}\,{\frac{{v^{\prime }(r)}}r}-{\frac{{1-u(r)}}{{r^2}}}
&=&{}+{\frac{{u(r)\,\phi {{^{\prime }(r)}^2}}}2}-{\frac{\phi {{{(r)}^2}}}2}
\\
\phi (r)-u(r)\,\phi ^{\prime \prime }(r) &=&+\phi ^{\prime }(r)\,\left( {%
\frac{(u(r)+1)}{r\ u(r)}}-{\frac{r\phi {(r)}^2}{2\ u(r)}}\right)
\end{eqnarray}

It can be observed that equations (4) and (6) are closed because they do not
depend on ${\it v}(r)$. In place of ${\it v}(r),$ below it will be sometimes
used the variable

\[
\nu (r)=\log \,({\it v}(r)). 
\]

The set of equations (4)-(6) have solutions showing a leading asymptotic
behavior near the origin $r=0$ of the following form

\begin{eqnarray}
u(r) &=&1-{\frac{{{{\it \phi _0}^2}\,{r^2}}}6}+... \\
{\it v}(r) &=&1-{\frac{{{{\it \phi _0}^2}\,{r^2}}}6}+... \\
\phi (r) &=&{\it \phi _0}\,\left( 1+{\frac{{r^2}}6}\right) +...
\end{eqnarray}

The approximate solutions near $r=0$ can be numerically extended away from
the origin by taking as the initial conditions the values at small radial
distance of $u(r),{\it v}(r),\phi (r)$ and $\phi ^{\prime }(r)$. The
numerical results for $u,{\it v}$ and $\phi $ are shown in Figs. 1,2
respectively. They correspond to the value $\phi _0=4.5$ for the unique
independent parameter of the approximate solution (7)-(9). The initial
values for the fields and the first derivative of the scalar field necessary
for the numerical algorithm were calculated from (7)-(9) at the radial
position $r=0.003$.

Fig.1 shows a decreasing behavior of the function $u(r),$ which seemingly
approaches zero linearly at some radius $r_0=0.502416$. The picture for the
variation of ${\it v}(r)$, while similarly decreasing, tends to approach a
constant value differing from zero when $r$ approaches $r_0$. The scalar
field, on another hand, increases away from the origin and approaches a
constant value at $r_0$, but with a fast growing slope.

In order to better understand the properties of the solution near $r_0$, the
finding of asymptotic solutions in this region was considered. After
assuming that the function $u(r)$ is vanishing at some radial distance with
a linear dependence, the following leading behavior for a solution with this
properties was determined

\begin{eqnarray}
u(r) &=& (r_{0} \phi_{r_0}^{2}-2/r_{0}) (r_{0}-r)+... \\
\phi(r) &=& \phi_{r_0}- 2 \sqrt{r_0-r}/{\sqrt {r_0}}+...
\end{eqnarray}

Afterwards, the two parameters in (10) and (11) were determined by making
the asymptotic function $u(r)$ given by (8) and its first derivative to
coincide with the associated values in the numerical solution, at some
particular point staying very close to $r_0=0.502416$. The selected point
was $r=0.5024$. The parameter values obtained were

\begin{eqnarray}
r_0 &=&0.502416, \\
\phi _{r_0} &=&5.13513.
\end{eqnarray}

The appropriate character of the found asymptotic behavior in describing the
considered numerical solutions, is indicated by the plotting of the
numerical solutions for the scalar field and its radial derivative in common
with the asymptotic functions. The Figs. 3 and 4 show these pictures for the
scalar field and the derivative, respectively. After considering that only $%
u(r)$ have been used for determining the parameters, the coincidence for
scalar field properties became a satisfactory one.

\section{ Interior and Schwartzschild solutions}

The problem of matching space-time regions separated by a boundary is a
subtle one and moreover when the boundary is singular as it is in our
situation. In this section an argument will be given, indicating that the
global field distribution composed of the just discussed internal solution
and the external Schwartzschild one, could be considered as satisfying the
coupled Einstein and Klein-Gordon equations without the need of external
sources or forces. Clearly, as the fields solve the equations out of $r=r_0$%
, the support of such sources reduces to this surface.

The discussion will proceed as follows. First, the mentioned global solution
will be regularized in the neighborhood of $r=r_0$. The modified fields,
after substituted in the equations will determine sources for each one of
them in order for the equations to be satisfied. After that, it will be
argued that the regularization can be constructed in such a form that all
the sources vanish in the sense of the generalized functions~\cite
{vladimirov}.

The coupled set of equations with sources satisfied by the regularized
solutions can be written as

\begin{eqnarray}
{\frac{{{u^{\prime }}_\epsilon (r)}}r} &=&{\frac{{1-u_\epsilon (r)}}{{r^2}}}-%
{\frac{{u_\epsilon (r)\,{{{\phi ^{\prime }}_\epsilon (r)}^2}}}2}-{\frac{{{{%
\phi _\epsilon (r)}^2}}}2}-e_\epsilon (r) \\
{\frac{u_\epsilon (r)}r}{\frac \partial {\partial r}}{\nu}_\epsilon (r) &=&{%
\frac{1-u_\epsilon (r)}{r^2}}+{\frac{{u_\epsilon (r)\,{{{\phi ^{\prime }}%
_\epsilon (r)}^2}}}2}-{\frac{{{{\phi _\epsilon (r)}^2}}}2}-p_\epsilon (r) \\
\phi _\epsilon (r) -u_\epsilon (r)\,{\phi ^{\prime \prime }}_\epsilon (r)
&=& {\phi ^{\prime }}_\epsilon (r)\,\left( {\frac{(u_\epsilon (r)+1)}r}-{%
\frac{r{\phi _\epsilon (r)}^2}2}\right) -j_\epsilon (r)
\end{eqnarray}

\noindent 
where $\epsilon $ is the only parameter defining the regularization for the
three fields and $e_\epsilon (r),p_\epsilon (r)$ and $j_\epsilon (r)$ are
the external sources. It can be noticed that $e_\epsilon (r)$ and $%
p_\epsilon (r)$ are proportional to the energy density and pressure of a
macroscopic body respectively.

Consider the three $\epsilon $ dependent fields under discussion defined for
all values of $\,\;r$ in the following continuous way

\begin{eqnarray}
\phi_{\epsilon}(r) = \left\{ 
\begin{array}{ll}
\phi_{i}(r) & r< r^{*}-\epsilon/2 \\ 
-{\phi_{i}((r^{*}-\epsilon/2) / \epsilon)} (r-r^{*}-\epsilon/2) & 
r^{*}-\epsilon/2 < r < r^{*}+\epsilon/2 \\ 
0 & r> r^{*}+\epsilon/2
\end{array}
\right. \\
u_{\epsilon}(r)  =\left\{ 
\begin{array}{ll}
u_{i}(r) & r< r^{*}-\epsilon/2 \\ 
f_{\epsilon}(r) & r^{*}-\epsilon/2 < r < r^{*}+\epsilon/2 \\ 
u_{e}(r) & r> r^{*}+\epsilon/2
\end{array}
\right. \\
\nu_{\epsilon}(r)  = \left\{ 
\begin{array}{ll}
\nu_{i}(r) & r< r^{*}-\epsilon/2 \\ 
{((\nu_{e}(r^{*}+\epsilon/2)-\nu_{i}(r^{*}+\epsilon/2)) / \epsilon)} * &\\ 
(r-r^{*}-\epsilon/2)  + \nu_{e}(r^{*}+\epsilon/2) & r^{*}-\epsilon/2 < r <
r^{*}+\epsilon/2 \\ 
\nu_{e}(r) & r> r^{*}+\epsilon/2
\end{array}
\right.
\end{eqnarray}

\noindent
where the subindexes $(i)\ $and $(e)$ indicate the internal and external
solutions for the fields. The external fields are given by the
Schwartzschild ones

\begin{eqnarray}
u_e(r) &=&1-{\frac{r_0}r\ ,} \\
exp\,(\nu _e(r)) &=&1-{\frac{r_0}r\ ,} \\
\phi _e &=&0.
\end{eqnarray}

The parameter $r^{*}(\epsilon )$, which differs from horizon radius $r_0$,
will be chosen in order to force the external sources to vanish in the sense
of the generalized functions after removing the regularization, that is,
when $\epsilon \rightarrow 0$. Let us define the auxiliary parameters

\begin{eqnarray}
\epsilon _1(\epsilon ) &=&r^{*}(\epsilon )+{\frac \epsilon 2}-r_0, \\
\epsilon _2(\epsilon ) &=&\epsilon -\epsilon _1(\epsilon ),
\end{eqnarray}

\noindent
and consider first the Klein-Gordon equation (16). After substituting
expressions (17) - (19) in it and passing to the limit $\epsilon \rightarrow
0$, the following expression for the external sources $j_\epsilon (r)$ is
obtained

\begin{eqnarray}
lim_{\epsilon \rightarrow 0^{+}\ }j_\epsilon (r) &=&\delta
(r-r_0)\,lim_{\epsilon \rightarrow 0^{+}\ } \left({\frac{u_\epsilon
(r^{*}(\epsilon )-{\frac \epsilon 2})}\epsilon }-{\frac{u_\epsilon
(r^{*}(\epsilon )+{\frac \epsilon 2})}\epsilon }  + \right. \nonumber \\
&&\left. {\frac 1{r^{*}(\epsilon )}}-{\frac{r^{*}\phi _i^2(r^{*}(\epsilon
)-\epsilon /2)}3}\right)  \nonumber \\
&=&\delta (r-r_0)\,lim_{\epsilon \rightarrow 0^{+}}\ \left({\frac{u_i(r_0-{%
\epsilon _2})}\epsilon }-{\frac{u_e(r_0+{\epsilon _1})}\epsilon }\right.+  \nonumber
\\
&& \left.{\frac 1{r^{*}(\epsilon )}}-{\frac{r^{*}(\epsilon )\phi
_i^2(r^{*}(\epsilon )-\epsilon /2)}3} \right)  \nonumber \\
&=&\delta (r-r_0)\,lim_{\epsilon \rightarrow 0^{+}}\ \left( r_0\phi
_i^2(r_0^{-})-2/r_0){\frac{\epsilon _2}\epsilon }-{\frac 1{r_0}}{\frac{%
\epsilon _1}{\epsilon _2}}\right.+  \nonumber \\
&& \left.{\frac 1{r^{*}(\epsilon )}}-{\frac{r^{*}\phi _i^2(r^{*}(\epsilon
)-\epsilon /2)}3} \right)
\end{eqnarray}

The coefficient of the $\delta $-function in (25) can be reduced to vanish
after using $\epsilon _2(\epsilon )=\epsilon -\epsilon _1(\epsilon )$ and
choosing the dependence $r^{*}(\epsilon )$ and $\epsilon _1(\epsilon )$ as
follows

\begin{eqnarray}
\epsilon _1(\epsilon ) &=&{\frac{(2/3)r_0\phi _i^2(r_0^{-})-1/r_0}{r_0\phi
_i^2(r_0^{-})-1/r_0}}\,\epsilon \ , \\
r^{*}(\epsilon ) &=&r_0+\epsilon _1(\epsilon )-\epsilon /2\ .
\end{eqnarray}

Then, the Klein-Gordon can be satisfied in the sense of the generalized
functions by properly fixing the regularization.

In the case of the Einstein equations (14) and (15), the substitution of the
regularized field expressions (17) - (19) give for the associated sources
limit relations of the form

\begin{eqnarray}
lim_{\epsilon \rightarrow 0^{+}\ }e_\epsilon (r) &=&lim_{\epsilon
\rightarrow 0}\,\left( O_1(\epsilon )\right) \,\delta (r-r_0)\ , \\
lim_{\epsilon \rightarrow 0^{+}\ }p_\epsilon (r) &=&lim_{\epsilon
\rightarrow 0}\,\left( O_2(\epsilon )\right) \,\delta (r-r_0)\ ,\ 
\end{eqnarray}

\noindent 
 where the $O_1$ and $O_2$ functions of $\epsilon$ tend to vanish,
whenever, the regularizing function $f_{\epsilon}(r)$ associated to
 $ u_{\epsilon} (r)$ in (18), is chosen to satisfy the unique condition 
$$
lim_{\epsilon \rightarrow 0^{+}}\ {\frac 1{\epsilon ^2}}\left(
\int_{r^{*}(\epsilon )-\epsilon /2}^{r^{*}(\epsilon )+\epsilon
/2}d\,rf_\epsilon (r)\right) =0\ . 
$$

Therefore, the defined regularization allows to argue that the deformed
gravitational plus scalar field configuration requires stabilizing external
sources that vanish in the generalized functions sense,
 when the regularization is removed. A more
detailed investigation of the stability of this field configuration will be
considered elsewhere.

Finally, it is interesting to consider the general relation expressing $%
u(\rho )$ as an spatial integral over the energy-momentum tensor component $%
T_o^{\;o\;}$. After integrating relation (1), the result takes the form

\begin{eqnarray*}
u(\rho ) &=&1-{(}8\pi k/c^4)\left( \frac 1\rho \right) \int_o^\rho ({\frac{{%
u(\rho )\,}\varphi {{{^{\prime }(\rho )}^2}}}2}+{\frac{{\it m}^2\ \varphi {{{%
(\rho )}^2}}}2)\ }\rho ^2d\rho \\
&=&1-{(}8\pi k/c^4)\left( \frac 1\rho \right) \int_o^\rho T_o^{\;o\;}{\ }%
\rho ^2d\rho
\end{eqnarray*}

Then, when $\rho $ approaches $\rho _{0}=r_{0}/ m$ the following relation
arises\mathstrut

\mathstrut

$\hspace{1in}(\mathstrut 4\pi /c^2)\int_o^{\rho _0}T_o^{\;o\;}\rho ^2d\rho
=(c^2\rho _0)/(2k)=M,$

\mathstrut

\noindent 
where $M$ is the total mass associated to the external Schwartzschild
solution as follows from

$\hspace{1in}\hspace{-0.5in}\smallskip \smallskip \smallskip u_{e}(\rho )=1-\rho
_0/\ \rho =1-\left( 2kM/c^2\right) /\rho .$

Therefore, the proper mass of the internal scalar field distribution is just
the necessary one to furnish the mass of the black hole as observed far away
in the external region.

\section{\protect\smallskip Conclusions}

Physical arguments about the possibility for the existence of black holes
with non-trapping interior are given. The internal solution consists of an
interacting with gravity scalar field which is bounded at the horizon. On
the contrary, the derivative of the scalar field diverges at the border.
This interior solution, after extended to the external region with the
Schwartzschild space-time, is regularized in the neighborhood of the horizon
in a particular way. It allows to show that the external sources which are
needed for the regularized fields to solve the equations, tend to vanish in
the generalized functions sense after removing the regularization. Thus, the
studied configuration suggests an alternative equilibrium limit for the
collapse of matter described, not by dust particles, but by continuous
scalar field configurations. As all the particles are ultimately described
by fields, the possibility for the physical relevance of such solutions
seems not out of place. It can be also speculated that a kind of behavior of
the wave equations coupled with gravity can exists, in which the formation
of trapped surface could be rejected dynamically. These questions will be
considered elsewhere.

Finally, we would like to remark on other possibilities for the further
continuation of the work. A first task which is imagined consists in to
numerically solve time dependent solutions produced by a sudden
disappearance of the external sources associated to the regularized fields.
Such an study could give information about the stability of the singular
solution. The stability would be reflected by a tendency of the time
dependent fields to reproduce the singular configuration.

\section*{Acknowledgments}

We are grateful to Jorge Castineiras, Victor Villanueva and Iraida Cabrera who
have participated in the earlier stages of the work and also acknowledge the
helpful discussions with Augusto Gonzalez, Daniel Sudarsky, Raphael Sorkin,
Erik Verlinde, Jose Socorro, Juan Rosales and the members of the Group of
Theoretical Physics of the Instituto de Matem\'{a}tica, Cibern\'{e}tica y
F\'{\i }sica. The support of the TWAS through the Research Grant 93-120
RG/PHYS/LA is also deeply acknowledged.

\newpage

\section*{Figure Captions}

\bigskip

{\bf Fig.1} Graphical representation for the behavior of $u(r)$ and $v(r)$.
Note that while $u(r)$ tends to vanish at $r_0$ the $v(r)$ function tends to
a finite value. \newline
\newline
\noindent
{\bf Fig.2} Graphical representation for the behavior of the scalar field $%
\phi (r)$. The field intensity grows in approaching the radius $r_0$ and its
derivative has a fast rising behavior near $r_0$. \newline
\newline
\noindent
{\bf Fig.3} Graphical representation for the behavior of $\phi(r)$ and its
asymptotic expression (11) after determining its parameters through the
fitting of the numerical results for $u(r)$ and its derivative at $r=0.5024$%
. \newline
\newline
\noindent
{\bf Fig.4} Graphical representation for the behavior of $\phi^{\prime }(r)$
and the derivative of the asymptotic expression for $\phi (r)$ (11) after
determining its parameters through the fitting of the numerical results for $%
u(r)$ and its derivative at $r=0.5024$.


\newpage

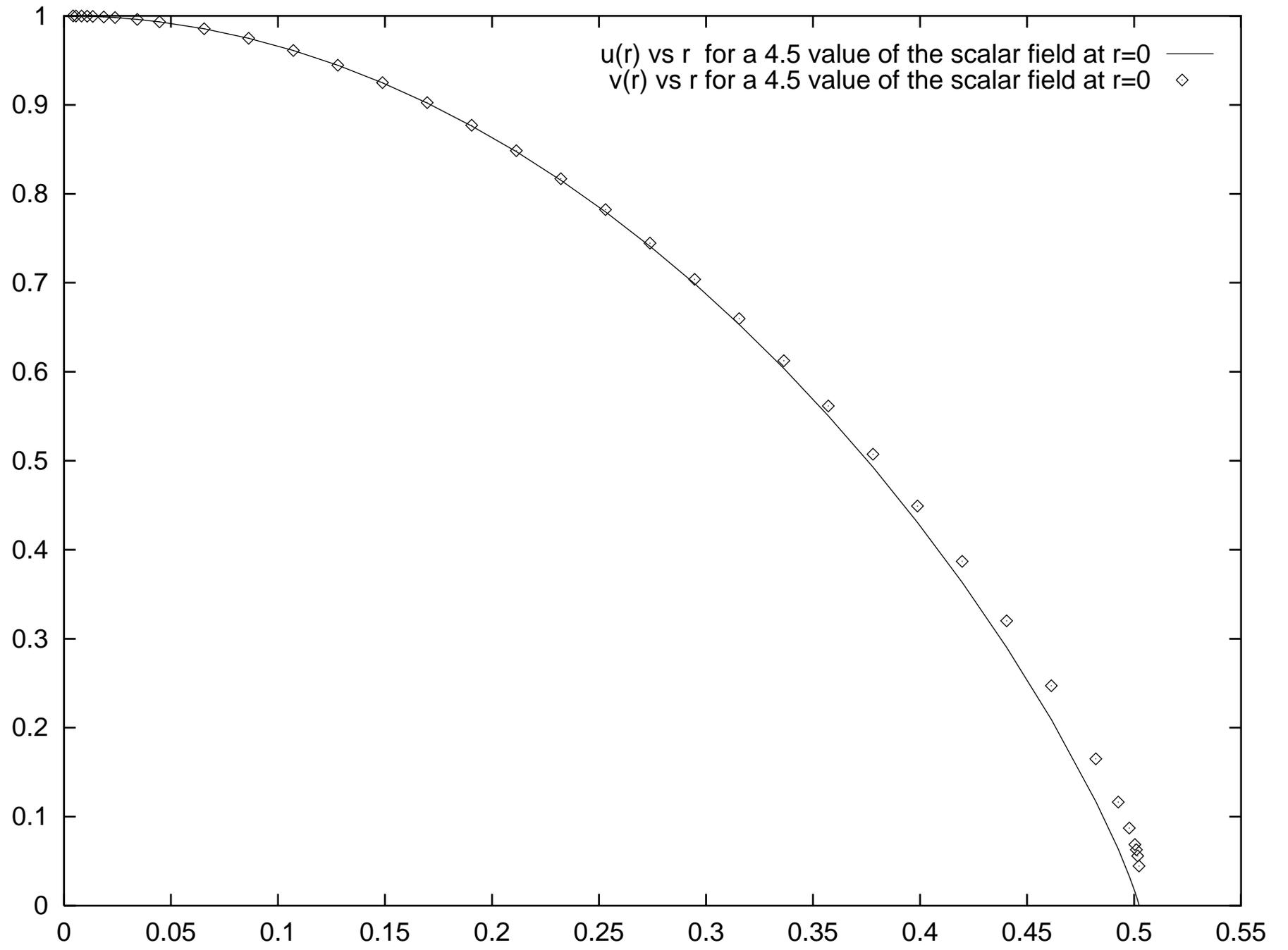
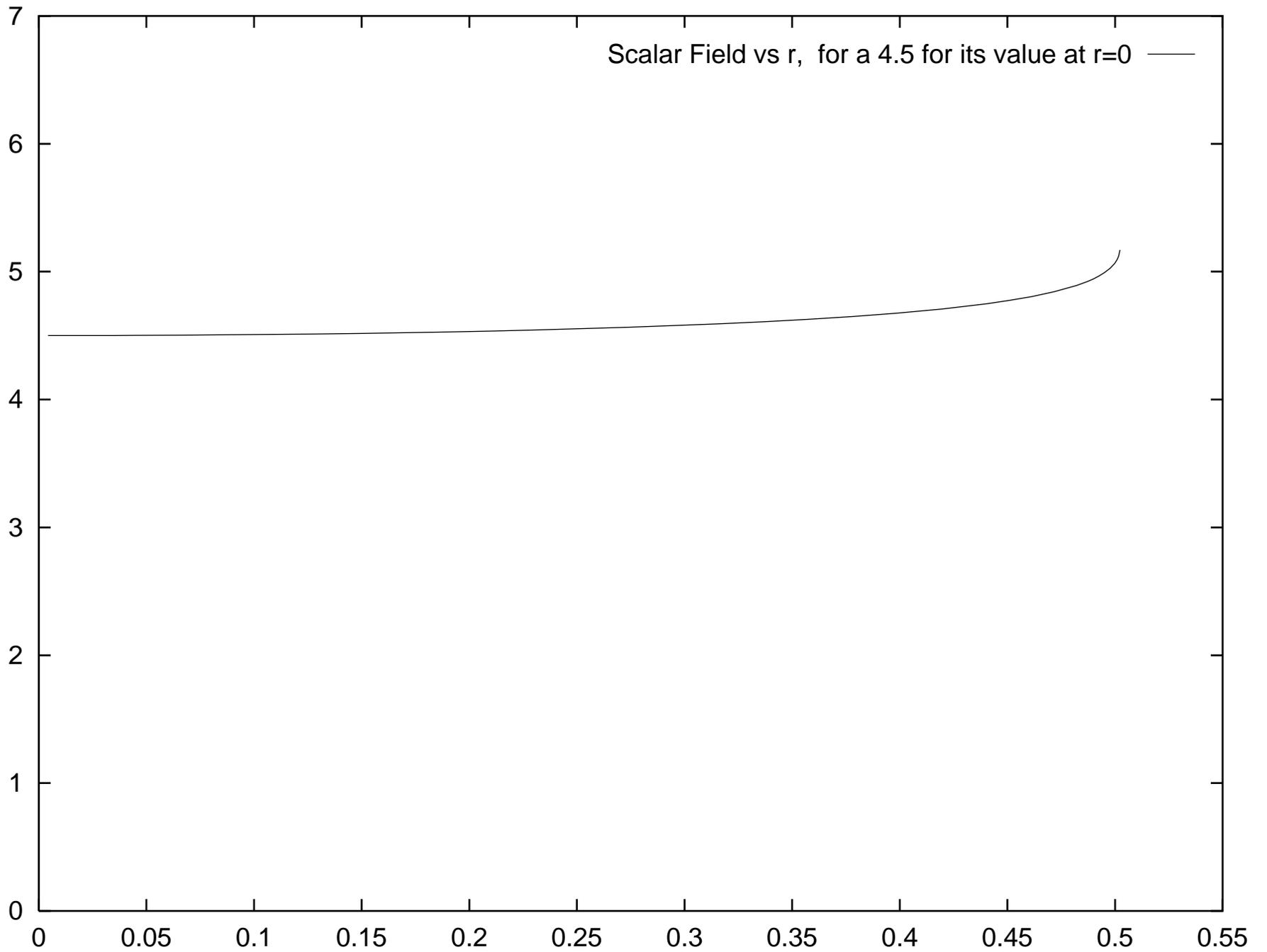
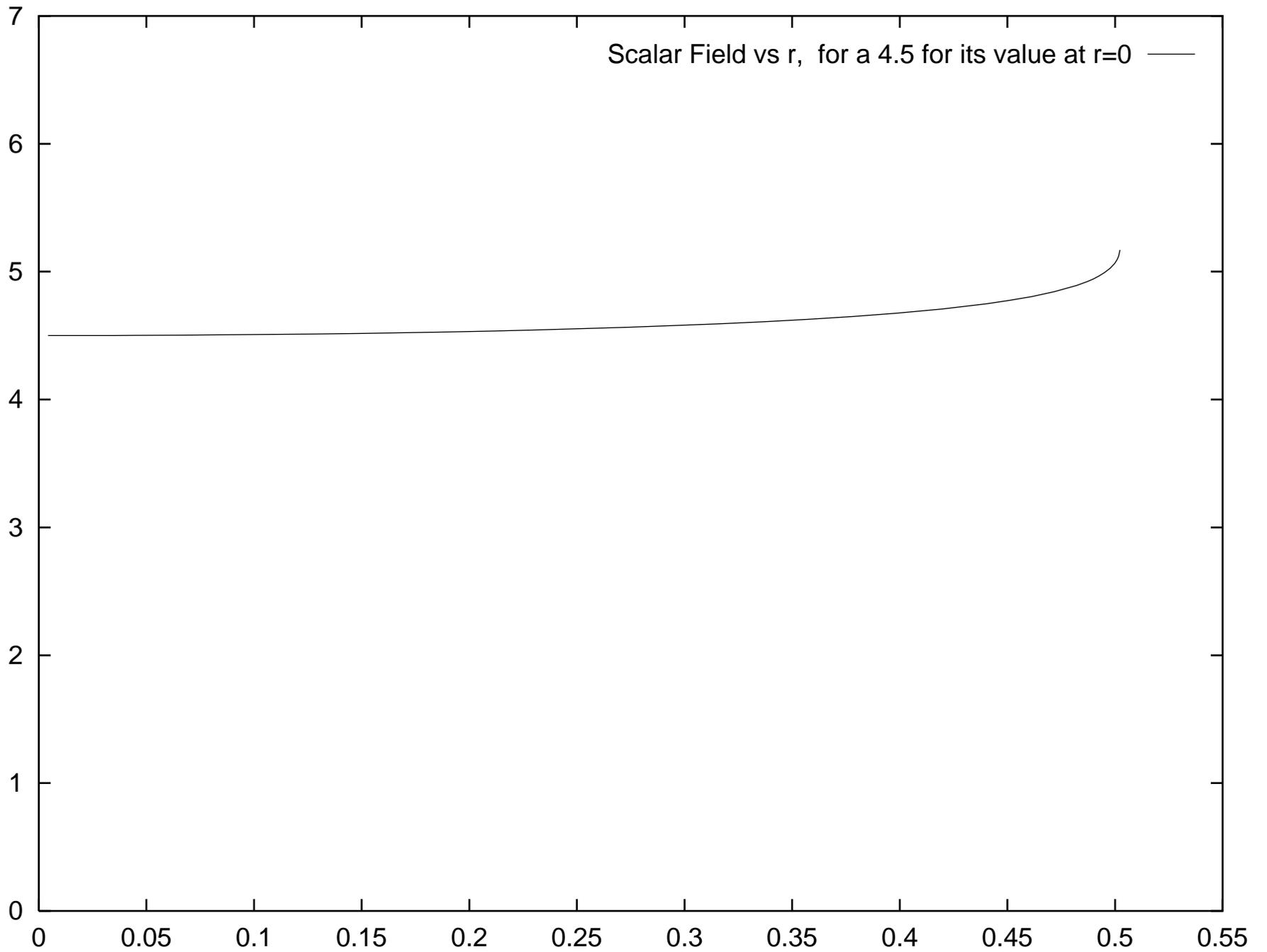
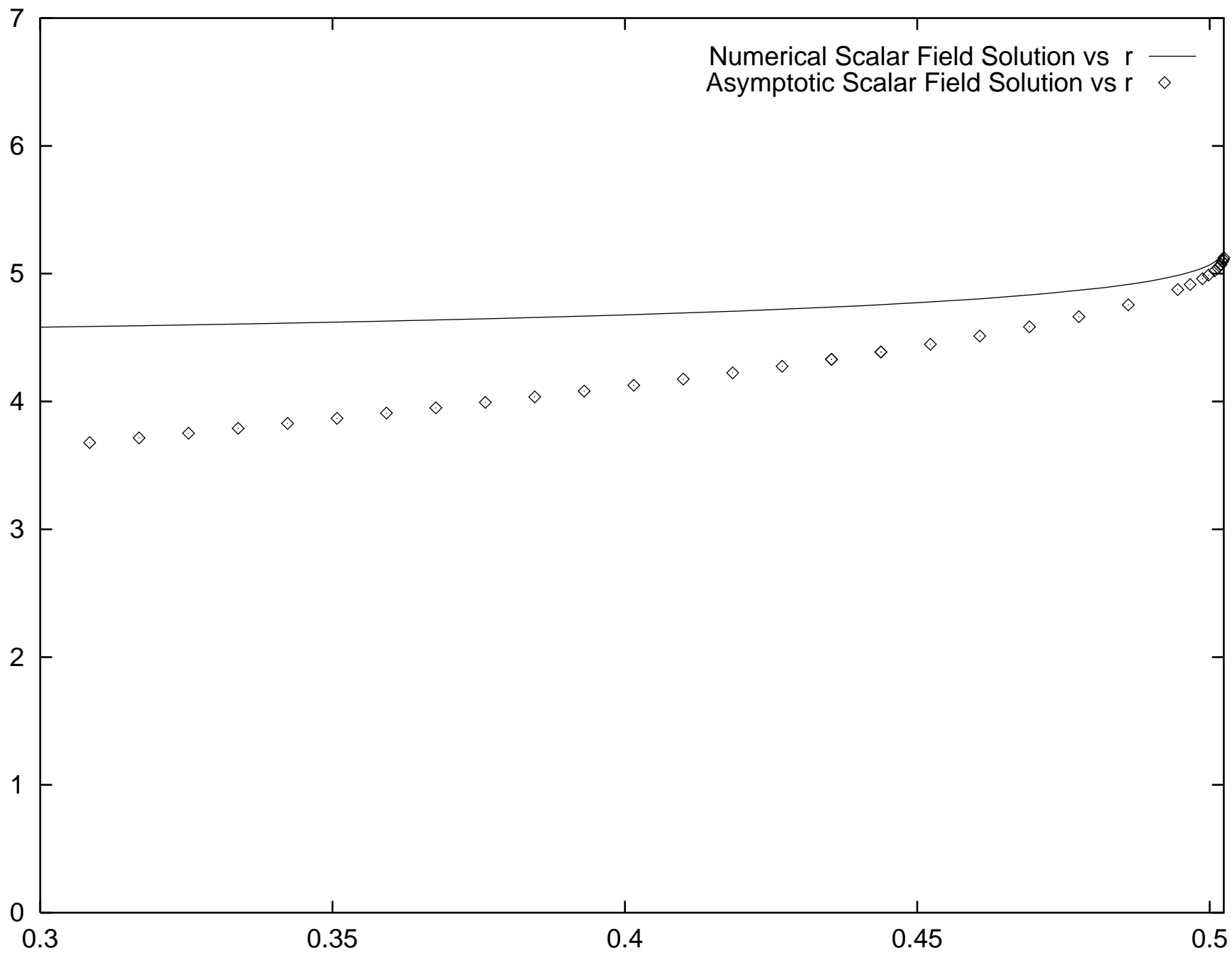
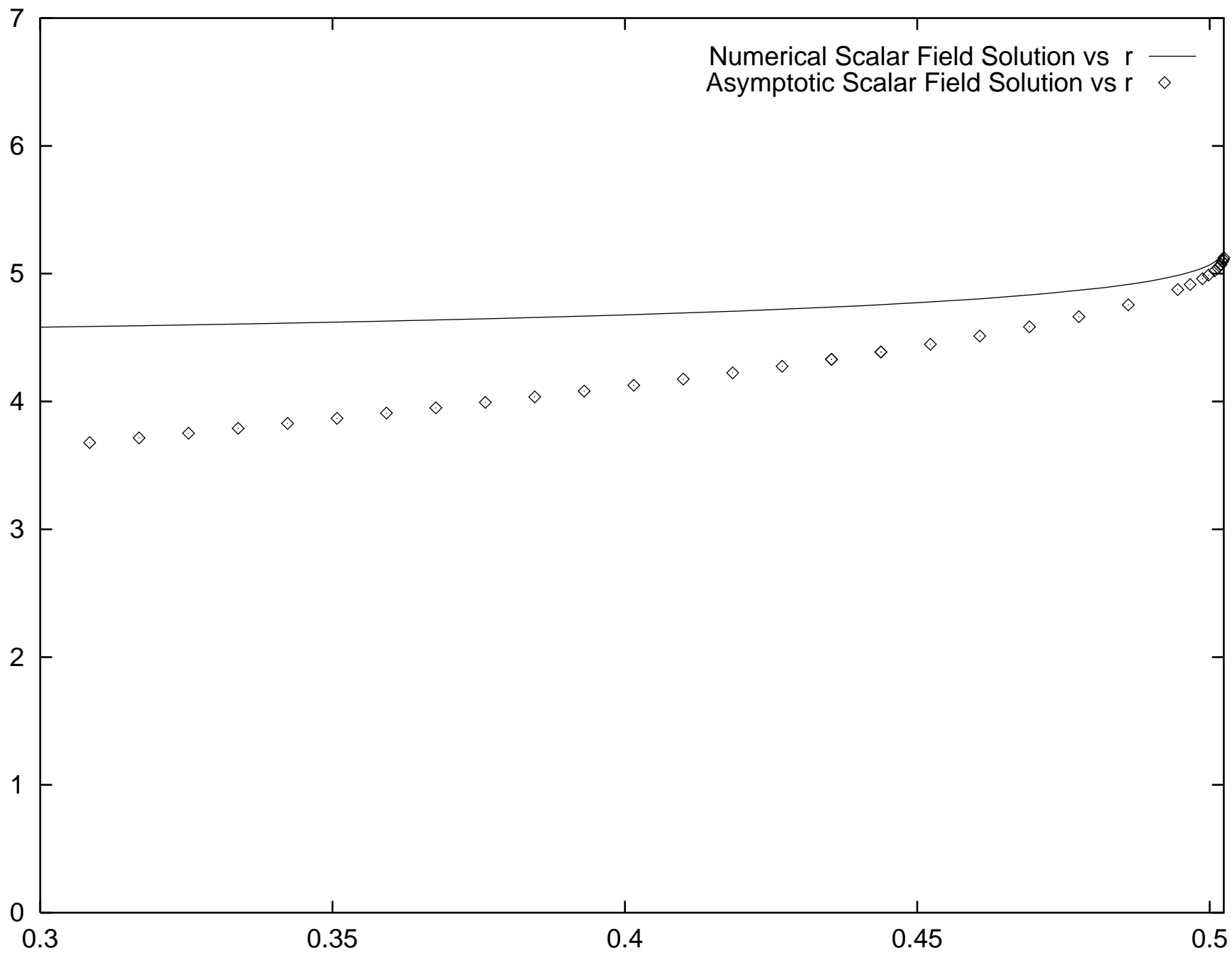
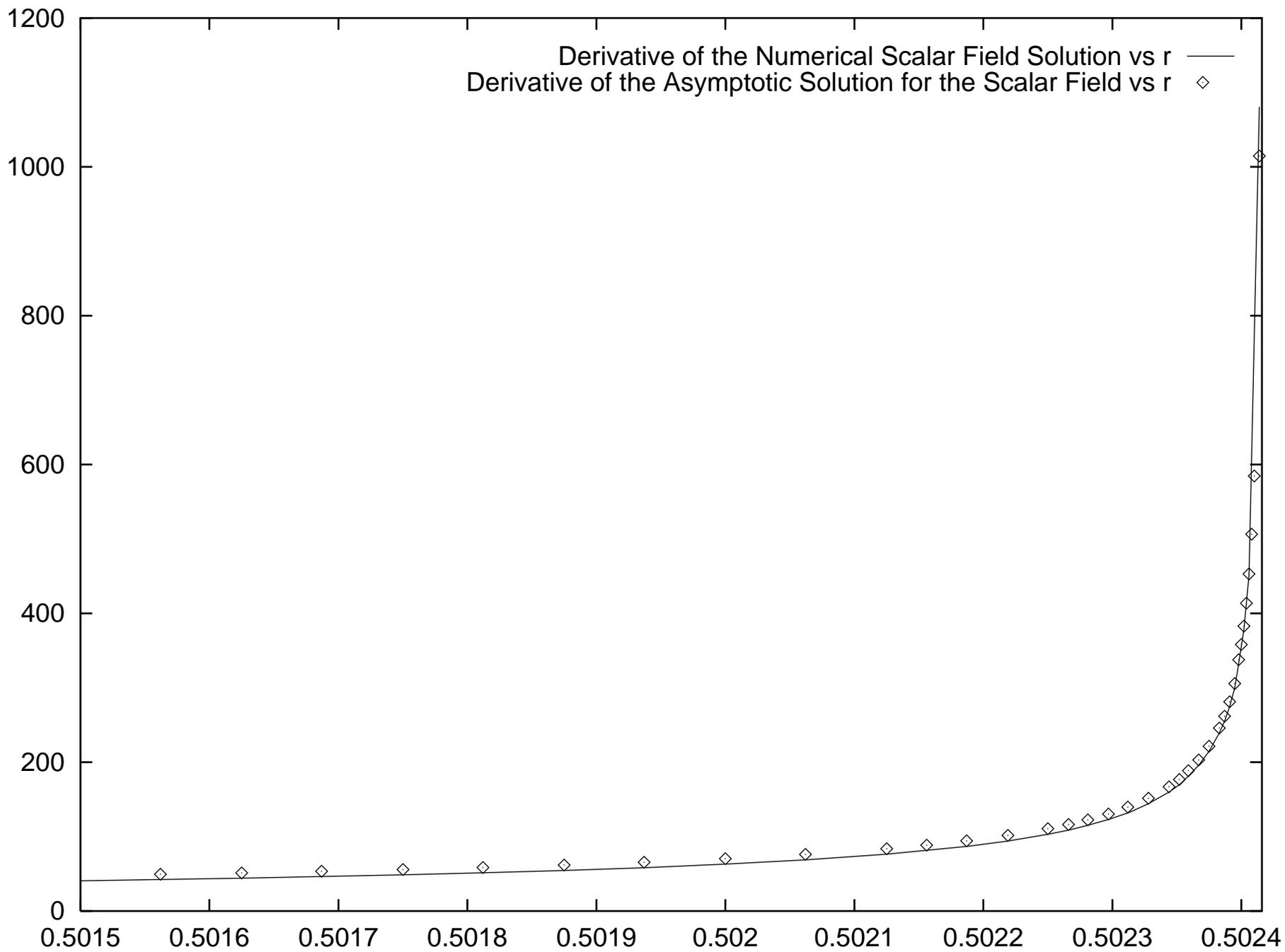
\end{document}